# Let's Burn a hole in the Binomial state of the radiation field

Priya Malpani[1,a)]

[1]*Indian Institute of Technology Jodhpur, Jodhpur 342037, India*

[a)]Corresponding author: malpani.1@iitj.ac.in

**Abstract.** In quantum optics, nonclassical properties of various quantum states of radiation field are frequently studied. Some of those states are finite dimensional and referred to as qudits. These states are important because of their potential applications in quantum information processing. Further, nonclassical states are those which do not have any classical counterpart. Consequently, to establish quantum supremacy, we always require nonclassical state. Recently, Sivakumar and Meher have studied the nonclassical properties of the number state filtered coherent state, and shown that the number state filtering introduces nonclassical features into coherent state which is otherwise classical. This observation motivated us to investigate the role of hole burning (state filtering) on a state which is already nonclassical. Specifically, we have selected a Binomial state which is known to be nonclassical as our test bed and burnt a hole at vacuum (equivalently filtered the vacuum state). To check the nonclassical properties of vacuum filtered binomial state, we have used Vogel's criterion, criterion of higher- and lower-order antibunching, criterion of higher-order sub-Poissonian photon statistics, Linear entropy etc. The investigation results show that vacuum filtered binomial state studied here is highly nonclassical, and the hole burning process enhances the nonclassical depth.

## INTRODUCTION

Various tasks which are impossible in the classical domain can be done in nonclassical regime [8]. Nonclassical states are used to achieve quantum supremacy. There are various nonclassical states, e.g. Photon added coherent state [1], Squeezed coherent state, Displaced Fock state [2], Photon added displaced Fock state, Photon subtracted displaced Fock state [3]etc. Recently, a new quantum state has been reported by Meher and Sivakumar known as Number state filtered coherent state [4]. The removal of any particular Number state from any quantum state is known as hole burning. In this article it is shown that how hole burning process increases the depth of nonclassicality.

## Quantum states of interest

Here the state of interest is an intermediate state known as binomial state [7] having hole at vacuum position.

### Binomial state with hole at vacuum

Binomial state is a combination of Number states in such a way that the distribution of occurrence probability of m photons is binomial. Binomial state is a finite dimensional state having analytical expression

$$|p,M\rangle = \sum_{n=0}^{M}\left[\frac{M!}{n!(M-n)!}p^n(1-p)^{M-n}\right]^{\frac{1}{2}}|n\rangle \qquad (1)$$

Vacuum filtered Binomial state can be obtained by simply eliminating vacuum $|0\rangle$ from Binomial state. An analytical expression for Number state filtered Binomial state and vacuum filtered Binomial state are given by Eq. (2) and (3) respectively.

$$|p,M,k\rangle = N_f \sum_{n=0,n\neq k}^{M}\left[\frac{M!}{n!(M-n)!}p^n(1-p)^{M-n}\right]^{\frac{1}{2}}|n\rangle \qquad (2)$$

$$|p,M,0\rangle = N_0 \sum_{n=0}^{M}\left[\frac{M!}{n!(M-n)!}p^n(1-p)^{M-n}\right]^{\frac{1}{2}}|n\rangle - N_0\left[(1-p)^M\right]^{\frac{1}{2}}|0\rangle \tag{3}$$

Here,

$$N_0 = \left[\sum_{n=0}^{M}\frac{M!}{n!(M-n)!}p^n(1-p)^{M-n} - (1-p)^M\right]^{-\frac{1}{2}}. \tag{4}$$

Here, $|n\rangle$ is normal Number state, p is real valued parameter whose value lies between 0 and 1. $N_0$ is the normalization constant for vacuum filtered Binomial state.

## Higher-Order moment

In case of vacuum filtered Binomial state or Binomial state having hole at vacuum position the analytical expression for higher-order moment can be given as

$$\langle \hat{a}^{\dagger t}\hat{a}^r\rangle = N_0^2 \sum_{n=1}^{M}\left[\frac{p^{2n-r+t}(1-p)^{2M-2n+r-t}}{(M-n+r-t)!(M-n)!}\right]^{\frac{1}{2}}\frac{M!}{(M-n)!}. \tag{5}$$

Putting the appropriate value of *t* and *r*, moment of any order can be obtained which is necessary for the calculation of nonclassicality witnesses reported here.

## Nonclassicality witnesses
## Lower- and higher-order antibunching

In this section, we study lower- and higher-order antibunching [7]. To do so, we use the following criterion of (*l* - *1*)th order antibunching [5]

$$d_f(l-1) = \langle \hat{a}^{\dagger l}\hat{a}^l\rangle - \langle \hat{a}^\dagger \hat{a}\rangle^l.$$

Signature of lower-order antibunching has been obtained as a special case for which *l* = 2. Here, it would be apt to note that for *l* >2 negative values of $d_f(l - 1)$ corresponds to higher-order antibunching of (*l* - *1*)th order. The idea to study HOA is to observe weaker nonclassicality. Here from the results it is clear that the nonclassicality is increasing as the order is increasing, the same is true for M.

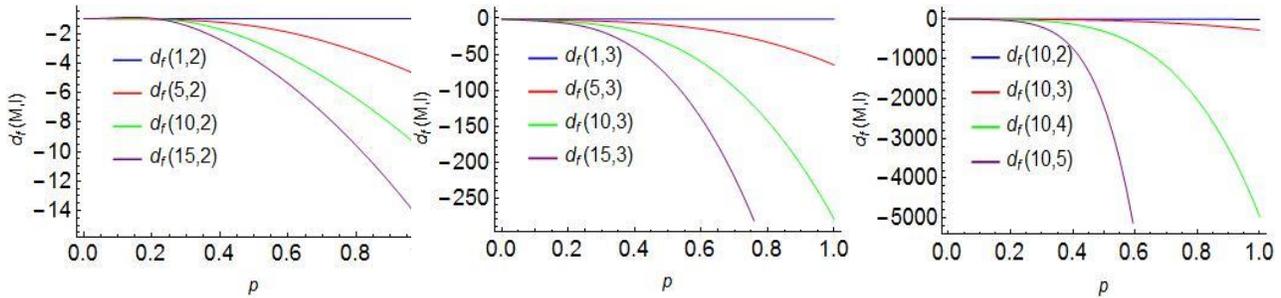

Figure 1: (Color online) lower and higher-order antibunching for vacuum filtered Binomial state as a function of probability (a) for different M and *l* = 2 (b) for different M and *l* = 3 and (c) for various order (*l*) for M = 10.

## Higher-order sub-Poissionian photon statistics

The study of higher-order nonclassicality can be done using Higher-order Sub Poissionian Photon Statistics. The analytical expression for the same is given as

$$D_h(l-1) = \sum_{e=0}^{l}\sum_{f=1}^{e} S_2(e,f)\binom{l}{e}(-1)^e d(f-1)\langle N\rangle^{l-e} < 0.$$

where $S_2(e,f)$ is the Stirling numbers of second kind, N is usual Number operator. As the order is increasing depth of nonclassicality is increasing, similar observation can be made for increase in M. For odd order the nonclassicalty is not observed which is consistent with previous results [3].

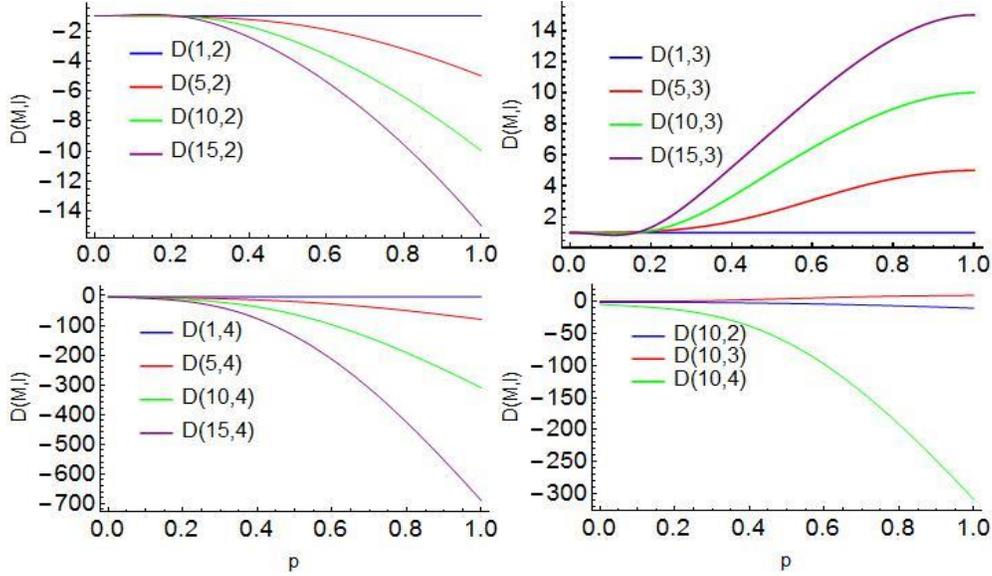

Figure 2: (Color online) higher-order sub-Poissionian photon statistics for vacuum filtered Binomial state as a function of probability (a) for different M and $l = 2$ (b) for different M and $l = 3$ (c) for different M and $l = 4$ and (d) for various order ($l$) and M = 10.

## Vogel's criterion

Vogel's nonclassicality [6] criterion in terms of moments can be expressed as

$$dv_n = \begin{bmatrix} 1 & \langle \hat{a} \rangle & \langle \hat{a}^\dagger \rangle & \cdots & \cdots \\ \langle \hat{a}^\dagger \rangle & \langle \hat{a}^\dagger \hat{a} \rangle & \cdots & \cdots & \cdots \\ \langle \hat{a} \rangle & \vdots & \vdots & \vdots & \vdots \end{bmatrix}$$

For nonclassical behaviour $dv_n < 0$. Form the results obtained here, nonclassicality is observed.

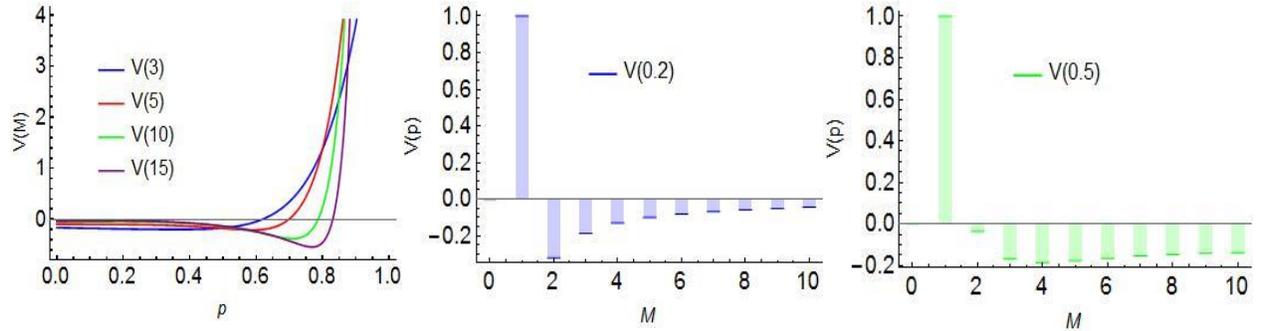

Figure 3: (Color online) Vogel's Criterion as a function of probability in (a), as a function of M for fixed probability in (b) and (c).

## CONCLUSION

In the concluding part, we would like to show the nonclassical behavior of vacuum filtered Binomial state using different higher and lower-order nonclassicality witnesses. The calculation and the results show that hole burning is the efficient way to achieve more nonclassicality.

## ACKNOWLEDGMENTS

P.M. wants to thank Dr. Anirban Pathak, Dr. Kishore Thapliyal and Dr. Nasir Alam for helpful discussions.